\documentclass[runningheads]{llncs}
\usepackage[T1]{fontenc}
\usepackage{graphicx}
\usepackage{color}
\usepackage{hyperref}

\usepackage{subfig}
\usepackage{algorithm}
\usepackage{algorithmicx,algpseudocode}
\usepackage{amsmath}
\usepackage{xcolor}
\usepackage{amsfonts}
\usepackage{subcaption}
\usepackage{multirow}
\usepackage{amssymb}
\usepackage{wrapfig}
\usepackage{marvosym}

\begin{document}
\title{FlowRec: Prior-Informed Flow Matching for Efficient Sequential Recommendation Generation}
\titlerunning{Flow Matching for Sequential Recommendation Generation}
\author{Li Li\inst{1} \and
Mingyue Cheng\inst{1}\textsuperscript{(\Letter)} \and Yuyang Ye\inst{2} \and Zhiding Liu\inst{1} \and Enhong Chen\inst{1} }
\authorrunning{L. Li et al.}
\institute{State Key Laboratory of Cognitive Intelligence, University of Science and Technology of China, Hefei, China\\
\email{\{lili0516,zhiding\}@mail.ustc.edu.cn, \{mycheng,cheneh\}@ustc.edu.cn}
\and
Rutgers University, New Brunswick, NJ, USA\\
\email{yuyang.ye@rutgers.edu}}
\maketitle              % typeset the header of the contribution
\begin{abstract}
Sequential recommendation aims to predict each user's next preferred item based on their historical interactions. Recently, diffusion-based approaches have demonstrated strong generative capability in modeling complex user preferences. However, they still face two inherent limitations: (i) Gaussian priors are misaligned with user-specific interests, and curved noise schedules lead to error accumulation and unstable training; (ii) the stochastic denoising process introduces additional randomness and substantial computational overhead. To address these issues, we propose FlowRec, a flow-matching-based framework that formulates preference evolution as continuous flows between personalized priors and target items. Specifically, FlowRec constructs an informative behavior-based prior distribution derived from users' historical interactions, offering a distributionally closer initialization to the target distribution. It then learns a vector field to guide straight preference flows toward target interests. Moreover, a single-step alignment objective with positive and negative samples further enhances semantic consistency between generated representations and ground-truth items. Finally, FlowRec adopts deterministic ODE-based generation, achieving efficient and stable inference. Extensive experiments on multiple benchmark datasets demonstrate that FlowRec consistently outperforms state-of-the-art baselines in both recommendation accuracy and inference efficiency. 

% Compared with diffusion models, flow matching offers greater flexibility in both the prior distributions and the sampling trajectories. 

\end{abstract}%
\section{Introduction}
\label{intro}

Sequential recommendation (SR) aims to model users' dynamic preferences from chronologically ordered historical interactions~\cite{hidasi2015session,tang2018personalized}. It has been widely deployed in various online platforms, such as e-commerce and streaming media services. Early studies relied on Markov chains and matrix factorization to capture short-term sequential dependencies~\cite{rendle2012bpr}, while recent deep learning-based methods significantly improve SR by modeling both long- and short-term user interests~\cite{tang2018personalized,cheng2024general}. More recently, diffusion models (DMs)~\cite{ho2020denoising,song2020score} have emerged as a powerful generative paradigm for SR, owing to their ability to model complex preference distributions. Typically, these models adopt a standard Gaussian distribution as the prior and treat the ground-truth item distribution as the target. By predefining a noise schedule that perturbs item embeddings with Gaussian noise in the forward process and iteratively
denoising them in the reverse process conditioned on user interaction sequences, diffusion-based recommenders~\cite{yang2023generate,li2023diffurec,DimeRec} achieve expressive preference generation.

% \vspace{-0.25in}
\begin{figure}[h]
    \centering
    \includegraphics[width=0.9\linewidth]{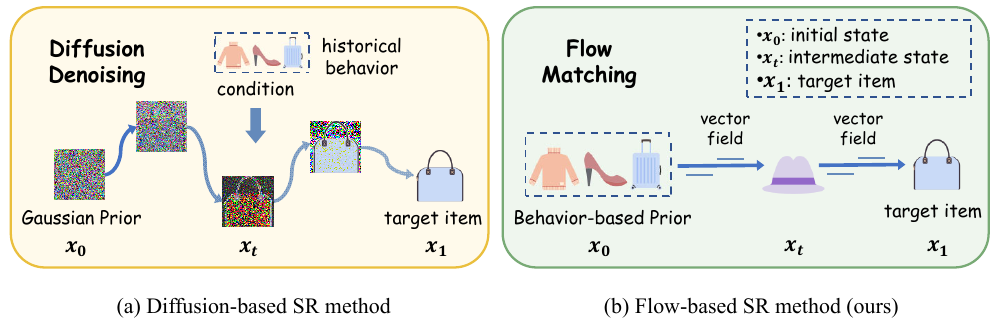}
    \caption{Comparison of our proposed method and diffusion-based method.}
    \label{fig:intro}
\end{figure}
% \vspace{-0.25in}

Despite their impressive effectiveness, diffusion-based approaches still suffer from two inherent limitations. First, their reliance on Gaussian priors and curved noise schedules limits their modeling capacity for SR. Gaussian priors introduce noise which is far from user-specific interests, forcing the model to reconstruct preferences from uninformative initializations. Although these models condition on users’ historical interactions, biased or insufficient contextual information may cause them to generate unrealistic items or drift away from users’ true preferences~\cite{xie2024bridging,liu2025flow}. Meanwhile, the curved noise schedules exacerbate error accumulation and noise sensitivity, leading to unstable and inaccurate predictions. Moreover, such schedules typically require many sampling steps during generation, further increasing computational cost. Second, the stochastic SDE sampler used during inference introduces additional noise and requires diffusion term computation, resulting in substantial computational overhead. Overall, these challenges highlight the need for a more stable and efficient generative paradigm that mitigates noise sensitivity and reduces redundant sampling during inference.

% most of them transform a simple prior—usually a standard Gaussian distribution—into a complex target distribution under rigid generation conditions (e.g. user's historical interaction sequence). However, since real-world recommendation data is often sparse and noisy, biased or insufficient conditions can easily lead to inaccurate recommendations. Second, the curved noise schedules adopted in these models are sensitive to noise perturbations, leading to error accumulation and unstable predictions. Moreover, such noise schedules often require many sampling steps for generation during the reverse process, substantially increasing computational overhead. Third, following the paradigm of image generation, existing diffusion-based SR methods mainly focus on denoising rather than explicitly modeling the dynamic evolution of user interests, thereby overlooking the intrinsic temporal dependencies essential for sequential recommendation. Overall, these challenges highlight the need for a more stable and efficient generative paradigm that mitigates noise sensitivity and reduces redundant sampling during inference.

To address the aforementioned challenges, we introduce the flow matching paradigm~\cite{lipman2022flow,liu2022flow}, a generalization of diffusion models that has shown strong generative capability across various domains, such as image~\cite{esser2024scaling}, speech~\cite{mehta2024matcha}, and text generation~\cite{hu2024flow}. Compared with diffusion models, flow matching offers greater flexibility in both the choice of prior distributions and the design of sampling schedules~\cite{tong2024improving,guerreiro2024layoutflow}. Leveraging these advantages, we propose {FlowRec}, a novel framework which explicitly models continuous preference flows from personalized priors to target items. The difference between FlowRec and diffusion-based methods is illustrated in Figure \ref{fig:intro}. Specifically, FlowRec encodes users' historical behaviors into latent intent representations, which serve as behavior-based priors, replacing the conventional Gaussian priors. These personalized priors provide informative and user-specific initialization which better approximates target items, thus reducing the learning difficulty compared to Gaussian priors. A learnable vector field is then optimized to construct straight preference trajectories, refining the initial representations toward the ground-truth items. Furthermore, to better adapt flow matching to SR, we introduce a single-step alignment loss that incorporates both positive and negative samples. This loss complements the flow-matching objective by semantically aligning the single-step estimation from the learned vector field with the target item representation. Finally, FlowRec adopts deterministic ODE-based generation, achieving efficient and stable inference. Extensive experiments on four public benchmark datasets demonstrate that FlowRec consistently outperforms state-of-the-art baselines in both recommendation accuracy and sampling efficiency.

% These predicted latent representations serve as the closest accessible approximation to the target items. By utilizing these informative and personalized representations as the initial distribution, we naturally reduce the learning difficulty of predicting target items compared to a Gaussian prior. 

% Specifically, we first encode users' historical behaviors into latent intent representations, which serve as the initial states for preference transitions. Then, we sample intermediate states between the initial states and the target items, adding noise to them. With these intermediate states, FlowRec learns a vector field that generates continuous trajectories from users' current states to target item embeddings. Finally, to better align flow matching with the recommendation objective, we tailor a single-step alignment loss which incorporates both positive and negative samples, thereby improving sampling efficiency and generation quality. Extensive experimental results on four benchmark datasets demonstrate that FlowRec achieves significant improvements in recommendation quality and inference efficiency.
We summarize the contributions of this paper as follows:
\begin{itemize} 
    \item We propose {FlowRec}, a novel SR framework based on flow matching, which explicitly models straight preference trajectories from personalized priors to target items, enhancing recommendation accuracy and inference efficiency.
    \item We design two key components that tailor flow matching to sequential recommendation: a behavior-based prior for personalized and informative initialization, and a single-step alignment loss that complements the flow-matching objective and improves semantic alignment.
    \item We conduct extensive experiments on four public benchmark datasets, demonstrating the superiority of FlowRec over a range of state-of-the-art baselines.
\end{itemize}

\section{Related Work}
\subsection{Sequential Recommendation}
Sequential recommendation aims to capture users' dynamic interests from historical interactions. Early methods model users' dynamic interest transitions using Markov chains, such as FMPC \cite{rendle2010factorizing} and Fossil \cite{he2016fusing}. In recent years, deep neural networks have become a dominant approach for SR, leading to notable performance improvements~\cite{kang2018self,liu2024llm,sun2019bert4rec}. GRU4Rec~\cite{hidasi2015session} employs the gated recurrent unit~\cite{cho2014properties} to predict the next item. Caser \cite{tang2018personalized} treats recent item sequence as an ``image'' and utilizes convolutional filters to capture sequential patterns. Self-attention-based methods further enhance the effectiveness of SR. SASRec~\cite{kang2018self} introduces uni-directional self-attention, while BERT4Rec~\cite{sun2019bert4rec} employs a bidirectional Transformer to encode user behavior sequences.

Beyond conventional methods, generative approaches such as VAEs \cite{kingma2013auto} and GANs \cite{goodfellow2014generative} have also been explored in SR. Methods like ACVAE \cite{ACVAE} and MFGAN~\cite{MFGAN} aim to model the uncertainty of user preferences and alleviate exposure bias. However, they often suffer from limited representation capacity and unstable training~\cite{liang2018variational}. Recently, diffusion models have emerged as a new promising paradigm in SR, achieving state-of-the-art performance~\cite{li2023diffurec,xie2024breaking,kim2024preference,cui2024context}. DiffuRec~\cite{li2023diffurec} and DiffRec~\cite{wang2023diffusion} directly apply diffusion models to SR. Specifically, these methods corrupt the next-item representation with Gaussian noise and iteratively reconstruct it through a denoising process. DreamRec~\cite{yang2023generate} utilizes a Transformer encoder to encode historical interactions as conditional guidance for the denoising procedure. CaDiRec \cite{cui2024context} employs a context-aware diffusion model to generate augmented views for SR via contrastive learning.

\subsection{Flow Matching}
Flow Matching~\cite{lipman2022flow,liu2022flow} (FM) is a recently proposed generative modeling paradigm based on continuous normalizing flows~\cite{chen2018neural}. While diffusion models are formulated as stochastic differential equation (SDE), FM models are based on ordinary differential equation (ODE). Unlike diffusion models starting from a Gaussian prior, FM allows more flexible prior distributions \cite{guerreiro2024layoutflow}. Instead of iterative denoising, FM learns a probability flow by regressing a vector field between the prior and target distribution, significantly accelerating the sampling process and improving stability. FM has demonstrated superior performance across various generative tasks, such as image generation~\cite{esser2024scaling,tong2023conditional}, speech generation \cite{mehta2024matcha}, and reinforcement learning \cite{kim2024preference}. Recent studies have begun adapting flow matching to recommendation tasks. FlowCF~\cite{liu2025flow} introduces a discrete flow framework to address sparsity and implicit feedback in collaborative filtering, while FMRec~\cite{liu2025flowbase} incorporates straight flow trajectory and task-specific objectives for sequential recommendation. Nevertheless, the potential of flow matching to leverage behavior-informed priors for SR remains underexplored, offering new opportunities for research.

\section{Preliminaries}
\subsection{Problem Statement}
Given an item set $ \mathcal{I} = \{i_1, i_2, \ldots, i_{|\mathcal{I}|}\}$ and a user set $ \mathcal{U}=\{u_1, u_2, \ldots, u_{|\mathcal{U}|}\}$, we denote each behavior sequence of user $u \in \mathcal{U}$ as $\mathcal{S} = [i_1^u, i_2^u, \ldots, i_{n-1}^u]$, where $i_j^u \in \mathcal{I}$ is the $j$-th item interacted with by user $u$ and $n$ is the sequence length. The goal of SR is to predict the next item $i_{n}^u$ that the user is likely to interact with, based on historical behavior sequence $\mathcal{S}$. This can be formulated as: 
\begin{equation}
    \mathop{\arg\max}_{i_j\in \mathcal{I}} P(i_{n}^u=i_j|\mathcal{S}),
\end{equation}
\noindent where $P$ denotes the likelihood of item $i_j$ being the next item, based on the user's historical interaction sequence $\mathcal{S}$. 

\subsection{Flow Matching}
Flow matching~\cite{lipman2022flow,liu2022flow} is a class of generative models that learn a continuous transformation from a prior distribution $p_0$ to a target distribution $p_1$. Unlike other generative paradigms, flow matching allows the prior distribution $p_0$ to be arbitrary. It aims to learn matching flows, represented by vector field, between $p_0$ and $p_1$ via an ODE: $d\boldsymbol{x}_t=u(\boldsymbol{x}_t, t)dt$, where $u(\boldsymbol{x}_t, t)$ denotes a time-dependent vector field describing the probability flow over time.

\paragraph{Forward Process.}
Given data samples $\boldsymbol{x}_1 \sim p_1(\boldsymbol{x})$ and prior samples $\boldsymbol{x}_0 \sim p_0(\boldsymbol{x})$, an interpolated flow path can be constructed as
\begin{equation}
    \label{eq1}
    \boldsymbol{x}_t = a_t\boldsymbol{x}_0 + b_t\boldsymbol{x}_1, \quad t\in [0,1],
\end{equation}
where $a_t$ and $b_t$ are predefined interpolation schedules. The corresponding vector field is given by the time derivative:
\begin{equation}
    u_t(\boldsymbol{x}_t |\boldsymbol{x}_1) =\frac{d\boldsymbol{x}_t}{dt} = a_t^\prime \boldsymbol{x}_0 + b_t^\prime \boldsymbol{x}_1,
\end{equation}
where $u_t(\boldsymbol{x}_t |\boldsymbol{x}_1)$ is the conditional vector field. The marginal vector field is defined as $u(\boldsymbol{x}_t, t) \triangleq \mathbb{E}_{p_t(u_t|x_t)}[u_t]$. A neural network $v_\theta$ parameterized by $\theta$ is trained to fit the marginal vector field: $\mathcal{L}_{\text{FM}}(\theta)=\mathbb{E}_{t, p_t(\boldsymbol{x}_t)} \left \Vert v_\theta(\boldsymbol{x}_t, t)-u(\boldsymbol{x}_t, t) \right\Vert^2$. However, directly computing this loss function is intractable due to the marginalization, so the conditional flow matching loss is used instead:
\begin{equation}
    \label{eq4}
    \mathcal{L}_{\text{CFM}}(\theta)=\mathbb{E}_{t, \boldsymbol{x}_0, \boldsymbol{x}_1} \left \Vert v_\theta(\boldsymbol{x}_t, t)-u_t(\boldsymbol{x}_t|\boldsymbol{x}_1) \right\Vert^2.
\end{equation}
Minimizing $\mathcal{L}_{\text{CFM}}$ is equivalent to minimizing $\mathcal{L}_{\text{FM}}$ with respect to $\theta$.

\paragraph{Reverse Process.}
After the vector field $v_\theta$is learned, new samples can be generated by integrating the ODE backward from the source distribution $p_0$:
\begin{equation}
    d\boldsymbol{x}_t = v_\theta(\boldsymbol{x}_t, t)dt, \quad \boldsymbol{x}_0 \sim p_0.
\end{equation}
Integrating from $t=0$ to $t=1$ yields deterministic trajectories toward the target distribution $p_1$. This reverse process defines a stable and efficient generative mechanism, in contrast to the stochastic sampling used in diffusion models.

\section{The Proposed FlowRec}
\subsection{Overview of FlowRec}
\begin{figure}
    \centering
    \includegraphics[width=.9\linewidth]{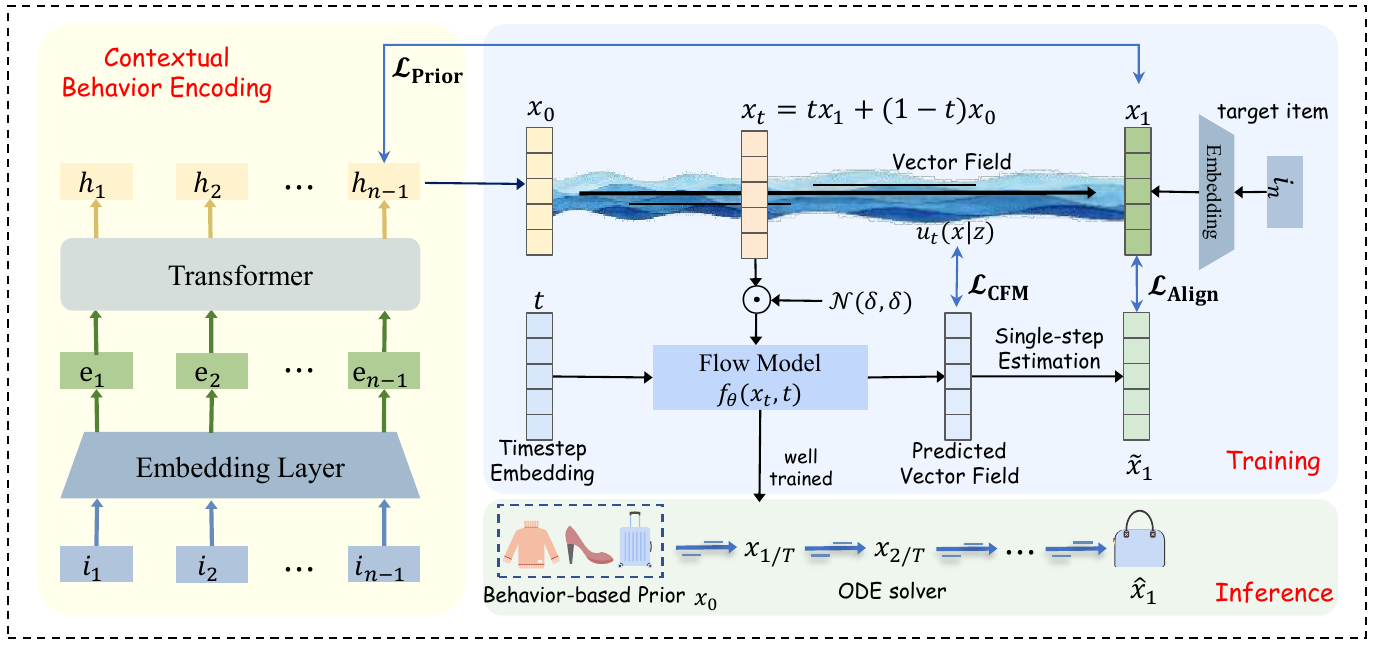}
    % \caption{Overview of FlowRec. The figure on the left is a Transformer-based behavior encoder. The two figures on the right illustrate the training phase and the inference phase, respectively.}
    \caption{ Overview of FlowRec. The left figure shows a Transformer-based behavior encoder, while the right figures illustrate the training and inference phases.}
    \label{overview}
\end{figure}

As illustrated in Figure~\ref{overview}, FlowRec learns a vector field to construct a straight flow path from a behavior-based prior to the target item. A user's interaction sequence is first encoded into a latent representation $\boldsymbol{x}_0 \sim p_0$, which serves as the closest accessible approximation to the target item. The embedding of the ground-truth target item is denoted as $\boldsymbol{x}_1 \sim p_1$. We train a flow model $f_\theta$ to estimate the vector field that guides the initial representation $\boldsymbol{x}_0$ toward the true target item $\boldsymbol{x}_1$. In addition to the flow matching objective, we tailor a single-step alignment loss that leverages the estimated vector field to obtain approximation $\tilde{\boldsymbol{x}}_1$ and semantically align it with the target item. FlowRec jointly optimizes the behavior-based prior loss, flow matching loss, and alignment loss to generate representations consistent with user preferences. During inference, FlowRec performs deterministic generation by solving the learned ODE from $\boldsymbol{x}_0$, yielding a refined item embedding  for efficient candidate ranking.

% \vspace{-0.45in}

\subsection{Preference Flows with Flow Matching}
% \subsection{{Contextual Behavior Encoding}}
\subsubsection{Contextual Behavior Encoding.}
Building upon the flow matching formulation, we construct a personalized and informative prior that captures users’ historical preferences. Unlike diffusion-based models that adopt a simple Gaussian prior, FlowRec leverages flow matching's flexibility to define a behavior-based prior distribution. This prior serves as the initial state of the preference flow, providing a closer and more informative initialization for generation.

Formally, each item $i\in\mathcal{I}$ is represented by an embedding vector $\boldsymbol{e} \in \mathbb{R}^d$, where $d$ denotes the embedding dimension. Given a user's historical interaction sequence $[i_1, i_2, \ldots, i_{n-1}]$, we obtain its embedding representation $\boldsymbol{E}_{1:n-1}=[\boldsymbol{e}_1, \boldsymbol{e}_2, \ldots, \boldsymbol{e}_{n-1}] \in \mathbb{R}^{(n-1)\times d}$, while the ground-truth next item is represented by $\boldsymbol{e}_{n}$. Following previous works~\cite{yang2023generate,xie2024bridging}, we employ a Transformer encoder to capture complex sequential dependencies, formulated as:
% We take the last output $\boldsymbol{h}_{n-1}$ of the Transformer encoder as the user's current preference state, which can be formulated as: 
\begin{equation}
    \boldsymbol{H} = \text{Trm}(\boldsymbol{e}_1, \boldsymbol{e}_2, \ldots, \boldsymbol{e}_{n-1}),
\end{equation}
where $\boldsymbol{H}\in \mathbb{R}^{(n-1)\times d}$ denotes the hidden states of the Transformer encoder and the last state $\boldsymbol{h}_{n-1}=\boldsymbol{H}[-1] \in \mathbb{R}^d$ is used as the behavior-based prior $\boldsymbol{x}_0$. We regard the distribution of these learned representations as the prior distribution $p_0$. Compared to a Gaussian noise prior, this prior preserves contextual information and user-specific intent. Notably, while we employ a Transformer encoder to model sequential dependencies, FlowRec is model-agnostic and can be integrated with other sequential encoders. It further optimizes the encoder’s output representations, consistently improving recommendation accuracy.

To ensure that $\boldsymbol{x}_0$ faithfully reflects users' true preferences, the behavior encoder is trained with a cross-entropy loss function: 

\begin{equation}
    \label{eq:priorloss}
    \mathcal{L}_\text{Prior} =-\frac{1}{|\mathcal{U}|}\sum\limits_{u\in \mathcal{U}}\log \frac{\exp(\boldsymbol{x}_0\cdot \boldsymbol{e}_{n}^u)}{\sum_{i \in \mathcal{I}}\exp(\boldsymbol{x}_0\cdot \boldsymbol{e}_{i}^u)}. 
\end{equation}

This objective encourages $\boldsymbol{x}_0$ to be close to the embedding of the ground-truth next item, effectively guiding the learned prior toward the target distribution. By providing such an informative initialization, FlowRec reduces the difficulty of learning preference trajectories and improves both convergence and recommendation accuracy.

\vspace{-0.1in}
\subsubsection{Vector Field Construction.} FlowRec learns a continuous vector field to model the smooth transition from the behavior-based prior $\boldsymbol{x}_0$ to the target item $\boldsymbol{x}_1$. Compared to curved noise schedule paths applied in diffusion-based models, our framework explicitly models straight preference flows, enabling stable and efficient preference evolution. 

To this end, we train a flow model $f_\theta$ to estimate the vector field that guides the transformation between the prior and target states. The straight trajectory, characterized by interpolation coefficients $a_t = 1 - t$ and $b_t = t$, is widely adopted in various studies~\cite{lipman2022flow,tong2024improving} for its simplicity and computational efficiency. We randomly sample a time step $t \in [0,1]$ from a uniform distribution. Following Eq.\eqref{eq1}, the flow path $\boldsymbol{x}_t$ is obtained by linear interpolation between $\boldsymbol{x}_0$ and $\boldsymbol{x}_1$:
\begin{equation}
    \boldsymbol{x}_t = (1-t) \boldsymbol{x}_0+ t \boldsymbol{x}_1,
    \label{linear}
\end{equation}
which defines a straight preference trajectory between the prior and target states.

Following~\cite{li2023diffurec}, to improve robustness and avoid direct exposure of $\boldsymbol{x}_0$ and $\boldsymbol{x}_1$, we introduce a stochastic modulation term $\boldsymbol{\lambda}\in \mathbb{R}^d$, where each component $\lambda_i \sim \mathcal{N}(\delta, \delta)$. The modulated input ($\boldsymbol{\lambda}\odot \boldsymbol{x}_t$) and the embedded time step $\boldsymbol{t}$ are then fed into a two-layer MLP parameterized by $\theta$:
\begin{equation}
v_\theta(\boldsymbol{x}_t, \boldsymbol{t})=f_\theta(\boldsymbol{x}_t, t)=MLP( (\boldsymbol{\lambda} \odot 
 \boldsymbol{x}_t) \oplus \boldsymbol{t}),
\end{equation}
\noindent where $\odot$ and $\oplus$ denote element-wise multiplication and concatenation, respectively. The predicted vector field $f_\theta(\boldsymbol{x}_t, t)$ captures the instantaneous direction of preference evolution, which will later be integrated through an ODE solver during inference to generate deterministic and smooth trajectories.

\vspace{-0.05in}
\subsection{Model Optimization}
\subsubsection{Conditional Flow Matching Loss.}
Following the straight trajectory defined in Eq.~\eqref{linear}, the corresponding conditional vector field can be expressed as: 
\begin{equation}
    u_t(\boldsymbol{x}_t|\boldsymbol{x}_1) = \frac{d\boldsymbol{x}_t}{dt}= \boldsymbol{x}_1 - \boldsymbol{x}_0,
    \label{eq: vector field}
\end{equation}
which is constant along the flow path and independent of $t$. According to Eq.~\eqref{eq4}, the vector field neural network $v_\theta$ (implemented as $f_\theta$ in FlowRec) is trained to approximate the true field $u_t$ via the Conditional Flow Matching (CFM) loss:
% \begin{equation}
% \label{eq: CFMloss}
%     \begin{aligned}
%         \mathcal{L}_\text{CFM}= &\mathbb{E}_{t, \boldsymbol{x}_0, \boldsymbol{x}_1} \left \Vert v_\theta(\boldsymbol{x}_t, t)-u_t(\boldsymbol{x}_t|\boldsymbol{x}_1) \right\Vert^2 \\
%          = & \mathbb{E}_{t, \boldsymbol{x}_0, \boldsymbol{x}_1}||f_\theta(\boldsymbol{x}_t, t)-(\boldsymbol{x}_1-\boldsymbol{x}_0)||^2.
%     \end{aligned}
% \end{equation}
\begin{equation}
\label{eq: CFMloss}
        \mathcal{L}_\text{CFM} = \mathbb{E}_{t, \boldsymbol{x}_0, \boldsymbol{x}_1}||f_\theta(\boldsymbol{x}_t, t)-(\boldsymbol{x}_1-\boldsymbol{x}_0)||^2.
\end{equation}

This objective enforces geometric consistency between the predicted and true vector fields, guiding the model to learn a vector field that drives the straight flow from the behavior-based prior $\boldsymbol{x}_0$ toward the target item $\boldsymbol{x}_1$.

However, minimizing this Euclidean distance alone only ensures that the predicted embedding is close to its corresponding target, without enforcing that it is farther from other negative items. In recommendation, such absolute proximity is insufficient — the model must learn relative distances that reflect ranking preferences. Without this constraint, the learned representation may remain ambiguous with respect to other items, leading to suboptimal ranking accuracy.

\subsubsection{{Single-Step Alignment Loss.}}
To address the above limitation, we introduce a \textit{single-step alignment loss} that provides a ranking-aware contrastive signal to complement the flow-matching objective. Starting from the middle state in Eq.~\eqref{linear}, the terminal state of the trajectory can be expressed by integrating the vector field over time: $\boldsymbol{x}_1 = \boldsymbol{x}_t + \int_t^1 u_t(\boldsymbol{x}_t|\boldsymbol{x}_1)dt$. 
% \begin{equation}
%     \boldsymbol{x}_1 = \boldsymbol{x}_t + \int_t^1 u_t(\boldsymbol{x}_t|\boldsymbol{x}_1)dt
% \end{equation}
In practice, since the conditional vector field $u_t(\boldsymbol{x}_t|\boldsymbol{x}_1)$ is constant and time-invariant under the straight trajectory assumption, we can approximate the integration using a first-order Euler step:
\begin{equation}
    \tilde{\boldsymbol{x}}_1 = \boldsymbol{x}_t + (1-t)\times f_\theta(\boldsymbol{x}_t, t).
\end{equation}
where $f_\theta$ denotes the learned approximation of the true vector field $u_t$. Then, we align the estimated embedding $\tilde{\boldsymbol{x}}_1$ with the ground-truth target embedding $\boldsymbol{e}_n$ using a contrastive cross-entropy loss:
\begin{equation}
    \label{align_loss}
    \mathcal{L}_\text{Align} =-\frac{1}{|\mathcal{U}|}\sum\limits_{u\in \mathcal{U}}\log \frac{\exp(\tilde{\boldsymbol{x}}_1\cdot \boldsymbol{e}_{n}^u)}{\sum_{i \in \mathcal{I}}\exp(\tilde{\boldsymbol{x}}_1\cdot \boldsymbol{e}_{i}^u)}. 
\end{equation}

This auxiliary loss explicitly incorporates both positive and negative items,
thereby enforcing that the predicted representation is not only close to its target but also discriminative from irrelevant candidates. By combining the geometric supervision from $\mathcal{L}_\text{CFM}$ with the ranking-aware contrastive signal from $\mathcal{L}_\text{Align}$, FlowRec achieves semantically aligned and discriminative preference trajectories, enabling accurate ranking with minimal sampling steps during inference.
% 同时利用了正负样本，解决了CFM loss in Eq.~\eqref{eq: CFMloss}中只利用正样本的限制。并且使用一步估计也使得在推理时更少的采样步数能获得更准确的结果。
\vspace{-0.1in}
\subsubsection{Joint Optimization.}Finally, the overall objective of FlowRec integrates all learning signals into a unified ed-to-end optimization framework:
\begin{equation}
    \mathcal{L} = \mathcal{L}_\text{Prior}+\alpha \mathcal{L}_\text{CFM}+\beta \mathcal{L}_\text{Align},
    \label{losseq}
\end{equation}
\noindent where $\alpha$ and $\beta$ are hyperparameters that control the the relative contributions of the flow-matching and alignment loss.

% 在推理阶段，we can generate the oracle item by encoding user historical behavior sequence as initial point and solving the ODE describing the flow as defined in Eq 1 through 训练好的$f_\theta$预测vector field. In practice, we adopt a numerical ODE solver (e.g. the Euler solver) to simulate the flow from $\boldsymbol{x}_0$ to the predicted embedding $\hat{\boldsymbol{x}}_1$. Given a predefined number of steps $T$, the initial sample x0 is moved along the direction predicted by the network in an autoregressive manner. Therefore, the backward process can be denoted as
\vspace{-0.1in}
\subsection{Model Inference}
During inference, FlowRec generates the next-item representation $\hat{\boldsymbol{x}}_1$ through deterministic integration of the learned flow. Given a user's historical interaction sequence, the behavior encoder first produces an initial state $\boldsymbol{x}_0$, which serves as the starting point of the learned flow. Once the flow model $f_\theta$ has been trained, inference proceeds by solving the ODE that defines the flow trajectory:
\begin{equation}
    d\boldsymbol{x}_t = f_\theta(\boldsymbol{x}_t, t)dt, \quad \boldsymbol{x}_0 \sim p_0.
\end{equation}

Unlike diffusion-based methods that adopts stochastic denoising with additional noise injection and diffusion term computation, this ODE-based formulation only involves simple numerical integration, ensuring both stability and efficiency. To numerically solve the ODE, we adopt the Euler method. Given a predefined number of integration steps $T$, the flow is computed iteratively as:
\begin{equation}
    \boldsymbol{x}_{t + \frac{1}{T}} = \boldsymbol{x}_{t} + \frac{1}{T}f_\theta(\boldsymbol{x}_{t}, t),
\end{equation}
\noindent where $t = 0, \frac{1}{T}, \frac{2}{T}, \ldots, 1-\frac{1}{T}$. After $T$ steps, we obtain the predicted embedding $\hat{\boldsymbol{x}}_1$ which approximates the user's next item embedding. Finally, FlowRec computes relevance scores between $\hat{\boldsymbol{x}}_1$ and all candidate item embeddings via an inner product, and recommends the top-$K$ items with the highest scores. 

The deterministic inference process not only eliminates the stochastic uncertainty of diffusion-based sampling but also substantially reduces computational overhead, enabling fast and stable recommendation generation.

\section{Experiments}
\label{experiment}
\subsection{Experimental Settings}
\subsubsection{Datasets.} We evaluate the proposed method on four public real-world datasets. All datasets have been widely used in sequential recommendation.
\begin{itemize} 
    \item \emph{Amazon Beauty} and \emph{Amazon Toys} are two subsets of the Amazon dataset~\cite{mcauley2015image}, containing user reviews and ratings for beauty products and toys.
    \item \emph{ML-1M}~\cite{harper2015movielens} contains one million movie ratings from approximately 6000 users on 4000 movies, provided by the MovieLens platform.
    \item \emph{Yelp} is a well-known business review dataset collected by Yelp.
\end{itemize}

Following previous studies~\cite{kang2018self,li2023diffurec}, we remove inactive users and unpopular items with fewer than five interaction records. The detailed statistics of all datasets are summarized in Table~\ref{datasets_table}.
% These datasets differ in both sequence length and size, encompassing a wide range of real-world scenarios.
% The statistics of the four datasets are illustrated in Appendix \ref{statistics_dataset}.

% \vspace{-0.35in}
\begin{table}[ht]
\caption{The detailed description and statistics of datasets.}
  \label{datasets_table}
  \centering
  \renewcommand{\arraystretch}{1}
    \tabcolsep=0.5cm
  \resizebox{\columnwidth}{!}{
      \begin{tabular}{lrrrrr}
        \hline
        Datasets & \#Sequence & \#Items & \#Actions & Avg\_len & Sparsity   \\
        \hline
        Beauty & 22,363  & 12,101 & 198,502 & 8.53 & 99.93\%     \\
        Toys & 19,412  & 11,924 & 167,597 & 8.63 & 99.93\%     \\
        ML-1M & 6,040  & 3,416 & 999,611 & 165.50 & 95.16\%     \\
        Yelp & 30,431  & 20,033 & 316,354 & 10.4 & 99.95\% \\
        \hline
      \end{tabular}
  }
\end{table}
% \vspace{-0.45in}

\subsubsection{Evaluation Settings.}
Following previous work~\cite{sun2019bert4rec,wang2023diffusion}, we set the max sequence length to 50 for all datasets. Recommendation performance is evaluated with two widely recognized metrics: HR@$k$ (Hit Rate) and NDCG@$k$ (Normalized Discounted Cumulative Gain), with results reported for $k =  5,10$. All datasets are split into training, validation, and test sets following the leave-one-out strategy. To ensure a fair comparison, we analyze ranking results on the full item set.

\subsubsection{Baseline Methods.}
To ensure a comprehensive assessment, we compare FlowRec with nine baseline methods in SR, divided into three categories: (i)~\textbf{classical methods}: GRU4Rec~\cite{hidasi2015session},  SASRec~\cite{kang2018self}, and Bert4Rec~\cite{sun2019bert4rec}, which model sequential dependencies through recurrent and self-attentive architectures; (ii)~\textbf{contrastive learning-based methods:} CL4SRec~\cite{xie2022contrastive} and CaDiRec~\cite{cui2024context}, which enhance representation robustness via jointly self-supervised contrastive objectives; (iii)~\textbf{generative-based methods:} ACVAE~\cite{ACVAE}, DiffuRec~\cite{wang2023diffusion}, and PreferDiff~\cite{liu2024preference}, which formulate recommendation as a generative modeling problem to capture the uncertainty of user preferences.

\subsubsection{Implementation Details.}
All experiments are conducted on a single NVIDIA GeForce RTX 4090 GPU. Baselines are implemented based on their official open-source codes. We employ a 4-layer bidirectional Transformer with 4 attention heads as the behavior-based prior encoder, and a 2-layer MLP as the flow model. The embedding size and hidden state size are both set to 128, with a batch size of 512. Dropout rates for the embedding and hidden layers are 0.1 and 0.3, respectively. The mean and variance $\delta$ of the Gaussian distribution are set to 0.001. We explore the sampling steps in $[1, 5, 10, 15, 20, 25, 30, 35]$. The model is optimized using Adam optimizer with an initial learning rate of 0.005. The loss function coefficients $\alpha$ and $\beta$ are tuned from $[5, 10, 15, 20]$ and $[1, 2, 3, 4, 5]$, respectively. Early stopping is applied with a patience of 10 epochs.

\vspace{-0.1in}
\begin{table}
  \centering
  \caption{Experimental results (\%) on the four datasets. We use ``H'' and ``N'' to represent HR and NDCG, respectively. The highest score is typed in bold, while the second-best score is underlined.}
  \label{performance}
  % \renewcommand{\arraystretch}{1.00}
  % \small
  \renewcommand{\arraystretch}{1}
    \tabcolsep=0.07cm
  \resizebox{\columnwidth}{!}{
      \begin{tabular}{c|l|ccc|cc|ccc|c|c}
        \hline
        Datasets & Metrics & GRU4Rec & SASRec & BERT4Rec & CL4SRec & CaDiRec & ACVAE & PreferDiff & DiffuRec  & \textbf{FlowRec} & Improv.  \\
        \hline
        \multirow{4}{*}{\rotatebox[origin=c]{90}{Beauty}} & H@5 & 3.1910 & 3.4623 & 2.4427 & 4.3438 & 4.9221 & 5.0912 & 4.6147 & \underline{5.3812} & \textbf{5.9123} & 9.87\% \\
        & N@5 & 2.1850 & 2.4802 & 1.5684 & 2.8437 & 3.1462 & 4.0230 & 3.4595 & \underline{3.9230} & \textbf{4.2166} & 7.48\% \\
        & H@10 & 4.7054 & 4.8909 & 4.0865 & 6.5313 & 7.2445 & 7.1074 & 5.8087 & \underline{7.5912} & \textbf{8.2760} &  9.02\% \\
        & N@10 & 2.6735 & 2.9359 & 2.9080 & 3.5489 & 3.9183 & 4.6807 & 3.8435 & \underline{4.6324} & \textbf{4.9804} &  7.51\% \\
        % & H@20 & 6.7746 & 5.3301 & 6.8885 & 6.5542 & 9.4128 & 10.2758 & 9.7907 & 6.9668 & \underline{10.797} & \textbf{11.3024} &  4.68\% \\
        % & N@20 & 3.1944 & 1.6371 & 3.4377 & 2.7179 & 4.2739 & 4.7245 & 5.4378 & 4.1359 & \underline{5.4428} & \textbf{5.7393} &  5.45\% \\
        \hline
        \multirow{4}{*}{\rotatebox[origin=c]{90}{Toys}} & H@5 & 1.7780 & 4.7127 & 2.2170 & 5.4039 & 5.3263 & 5.5878 & 5.0421 & \underline{5.7039} & \textbf{6.2539} & 9.64\%\\
        & N@5 & 1.2350 & 3.4509 & 1.3413 & 3.7329 & 3.6209 & 4.2412 & 3.8068 & \underline{4.2871} &\textbf{4.6209} & 7.79\%\\
        & H@10 & 2.7398 & 6.4063 & 3.5463 & 7.4232 & \underline{7.6914} & 7.3213 & 6.0890 & 7.3512  & \textbf{8.2261} & 6.95\% \\
        & N@10 & 1.5402 & 3.9944 & 1.7706 & 4.3782 & 4.3842 & \underline{4.9891} & 4.1855 & 4.8195 & \textbf{5.2578} &  5.39\% \\
        % & H@20 & 3.9669 & 4.3336 & 8.5108 & 5.5532 & 10.2565 & \underline{10.6320} & 9.6235 & 7.0265 & 9.6790 & \textbf{10.8982} & 2.50\% \\
        % & N@20 & 1.8504 & 1.4216 & 4.5258 & 2.2702 & 5.0895 & 5.1246 & \underline{5.6199} & 4.4238 & 5.4065 & \textbf{5.9304} & 5.53\%\\
        \hline 
        \multirow{4}{*}{\rotatebox[origin=c]{90}{Yelp}} & H@5 & 1.6976 & 2.6009 & 2.6602 & 1.9684 & 2.3183 & 1.3223 & 2.0505 & \underline{3.8268} &\textbf{3.9683} & 3.70\%\\
        & N@5 & 1.1749  & 1.8038 & 1.8953 & 1.2151 & 1.4708 & 0.9135 & 1.4777 & \underline{2.6579} & \textbf{2.7929} & 5.08\%\\
        & H@10 & 2.6390 & 3.9924 & 4.2884 & 3.4274 & 3.8524 & 2.3951 & 2.7800 & \underline{5.7047} & \textbf{5.9408} & 4.14\%\\
        & N@10 & 1.4750 & 2.2499 & 2.4194 & 1.6846 & 1.9625 & 1.2754 & 1.7135 & \underline{3.2619} & \textbf{3.4253} &  5.01\%\\
        % & H@20 & 4.3206 & 6.5893 & 6.1606 & 6.4788 & 5.8164 & 6.0241 & 4.3112 & 3.4701 & \underline{8.6867} & \textbf{8.7955} &  1.25\%\\
        % & N@20 & 1.8961 & 3.4631 & 2.7940 & 2.9681 & 2.2848 & 2.5195 & 1.808 & 1.8885 & \underline{4.0106} & \textbf{4.1419} &  3.27\%\\
        \hline
        \multirow{4}{*}{\rotatebox[origin=c]{90}{ML-1M}} & H@5 & 5.8405 & 9.0894 & 8.222 & 1.7219 & 12.7745 & 9.5543 & 7.7716 & \underline{15.2918} & \textbf{15.5848} & 1.92\%\\
        & N@5 & 3.8281 & 5.5502 & 5.1276 & 1.0221 & 8.3723 & 6.5349 & 5.0609 & \underline{10.0919} & \textbf{10.3634} &  2.69\%\\
        & H@10 & 9.2965& 15.5286 & 13.7086 & 3.3609 & 20.2898 & 16.1316 & 11.9801 & \underline{23.7949} &\textbf{24.1482} &  1.48\%\\
        & N@10 & 4.9321 & 7.6131 & 6.8960 & 1.5433 & 10.7855 & 7.8832 & 6.2955 & \underline{12.8233} & \textbf{13.1228} &  2.34\%\\
        % & H@20 & 15.6052 & 22.5507 & 24.2513 & 22.1137 & 6.2583 & 29.7184 & 25.6600 & 18.3112 & \underline{35.1100} & \textbf{35.8462} &  2.10\%\\
        % & N@20 & 6.5166 & 8.4304 & 9.8041 & 9.0102 & 2.2637 & 13.1591 & 12.0612 & 8.3767 & \underline{15.6776} & \textbf{16.0764} &  2.54\%\\
        \hline
      \end{tabular}
  }
\end{table}
\vspace{-0.25in}
\subsection{Overall Performance}
The overall performance of FlowRec and all baselines across four benchmark datasets is presented in Table~\ref{performance}. We can make the following observations: {(i) Generative models, especially diffusion models, outperform traditional methods, demonstrating their strong capability in modeling complex data distributions and preference uncertainty; {(ii) FlowRec achieves state-of-the-art results on all datasets.} For instance, on the Beauty dataset, it improves HR@5 and HR@10 by 9.87\% and 9.02\% over the strongest baseline DiffuRec. These results confirm the effectiveness and generalizability of our framework; {(iii) FlowRec benefits from its behavior-based prior and learnable vector field.} Unlike models such as SASRec that treat next-item prediction as a one-step classification, FlowRec explicitly models continuous preference transitions from a personalized prior to the target item. This formulation enables it to refine both the initial representation and the generated trajectory iteratively. Compared with diffusion-based methods (e.g., DiffuRec, PreferDiff), FlowRec generates straight and stable preference flows from informative priors, avoiding stochastic sampling noise and achieving deterministic, efficient inference.

\subsection{Generalizability on Insufficient Context}
\subsubsection{Head and Long-Tail Items.} Following previous work~\cite{cheng2024general}, we split the top 20\% most frequent items as head items and the rest as long-tail items. Users are grouped based on their most recent interaction $i_{n-1}$ before the target item. As shown in Figure~\ref{fig:popular}, long-tail items are typically more difficult to recommend due to their sparse interaction contexts. DiffuRec, which reverses a diffusion process from pure Gaussian noise, is sensitive to such weak context conditions, often producing unstable predictions. In contrast, FlowRec leverages informative and personalized priors to generate deterministic preference trajectories, ensuring stable inference even under sparse supervision. Consequently, FlowRec achieves significant gains on long-tail items. For example, on the Toys dataset, FlowRec achieves only an 8.85\% improvement over DiffuRec on head items but a substantial 21.28\% gain on long-tail items. A similar trend is observed across other dataset, demonstrating FlowRec's strong generalization ability.
% \vspace{-0.25in}
\begin{figure}
    \centering
    \includegraphics[width=\linewidth]{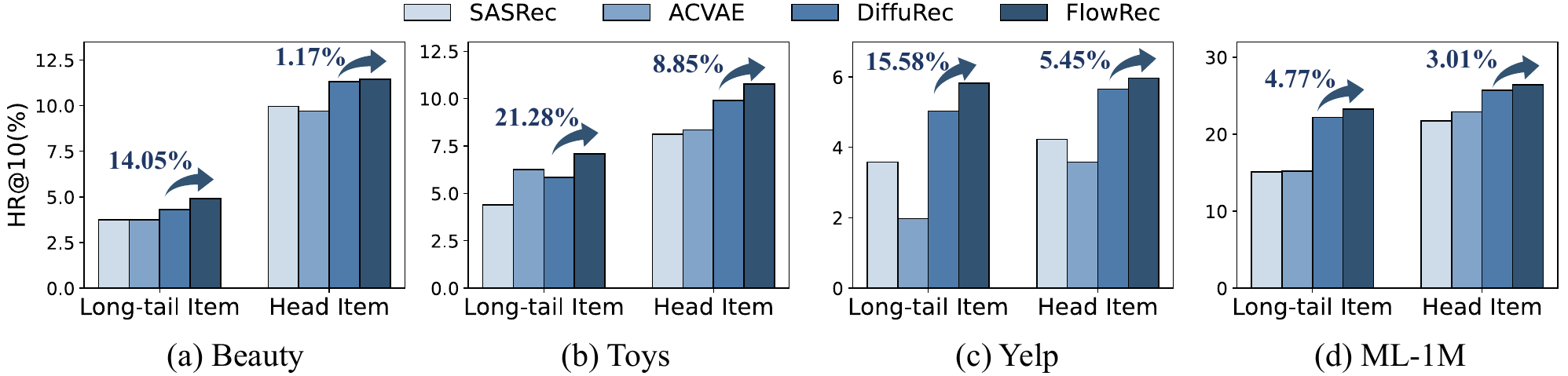}
    \caption{Performance on the long-tail and head items.}
    \label{fig:popular}
\end{figure}
\vspace{-0.35in}

\subsubsection{Different Sequence Length.}
Based on the percentile of sequence length, we divide all datasets into three groups (short, middle, and long) in a 1:2:1 ratio. The results are shown in Figure~\ref{fig:seqlen}. Short sequences often lack sufficient information to capture users' current preferences for a precise recommendation, whereas it is also challenging to handle very long sequences. Both DiffuRec and FlowRec perform  better on long sequences than SASRec and ACVAE, showing their superior ability to capture complex historical patterns. Notably, FlowRec achieves significantly higher accuracy on short sequences, benefiting from its deterministic ODE formulation that avoids noise injection and yields more stable inference.

% \vspace{-0.25in}
\begin{figure}
    \centering
    \includegraphics[width=\linewidth]{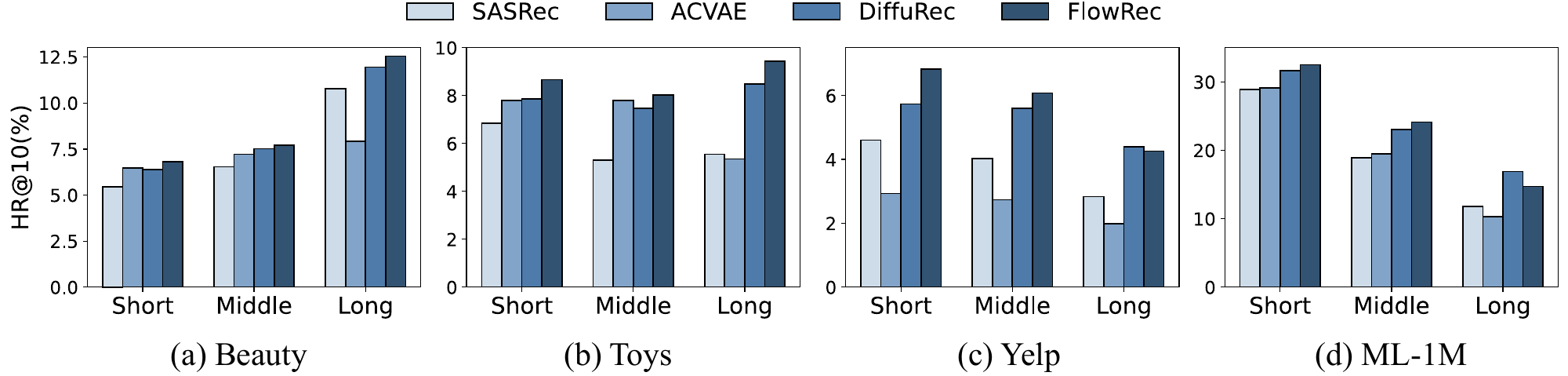}
    \caption{Performance on different length of sequence.}
    \label{fig:seqlen}
\end{figure}
\vspace{-0.35in}

\subsection{Ablation Study}
We conduct an ablation experiments to verify the effectiveness of each component in FlowRec. The experimental results are presented in Table~\ref{ablation}.

Removing any of the three objectives — $\mathcal{L}_{\text{Prior}}$, $\mathcal{L}_{\text{CFM}}$, or $\mathcal{L}_{\text{Align}}$ — leads to a clear performance degradation, verifying that all components are indispensable. In particular, eliminating $\mathcal{L}_{\text{Align}}$ causes a notable drop, highlighting the necessity of contrastive alignment for distinguishing relevant items from negatives. The variant without $\mathcal{L}_{\text{Prior}}$ performs worst, confirming that the behavior-based prior provides an informative and personalized initialization that simplifies the generative process.

We further replace the Transformer encoder with a GRU (``w GRU''). The performance decrease indicates that the quality of prior representations strongly influences overall generation. Nevertheless, this variant still surpasses GRU4Rec by a large margin, demonstrating that FlowRec’s flow-based formulation remains effective regardless of encoder choice.

Finally, we compare the recommendation accuracy of the initial state $\boldsymbol{x}_0$ and the generated state $\hat{\boldsymbol{x}}_1$. The strong performance of $\boldsymbol{x}_0$ verify that the learned priors are already well-aligned with user intent, while the full FlowRec further refines these representations through continuous flow modeling, achieving the highest accuracy and most stable convergence.

% \vspace{-0.35in}
\begin{table}[]
    \caption{Results(\%) of ablation experiments.}
    \label{ablation}
    \centering
    \renewcommand{\arraystretch}{1}
    \tabcolsep=0.2cm
    \resizebox{.97\columnwidth}{!}{
    \begin{tabular}{c|c|ccccc|c}
        \hline
         Datasets & Metrics & w/o $\mathcal{L}_{\text{Prior}}$ & w/o $\mathcal{L}_{\text{CFM}}$ & w/o $\mathcal{L}_{\text{Align}}$ & w GRU & Initial & FlowRec\\
         \hline
         \multirow{4}{*}{Beauty} & HR@5 & 2.9476 & 5.3070 & 3.7949 & 4.3986 & 5.6843 &\textbf{5.9125} \\
          & NDCG@5 & 1.9851 & 3.7331 & 2.4130 & 3.0822 & 4.0517 &\textbf{4.2166} \\
          & HR@10 & 4.4039 & 7.7036 & 5.8232 & 6.4790 & 8.0083 &\textbf{8.2760}\\
          & NDCG@10 & 2.4586 & 4.5058 & 3.0668 & 3.7474 & 4.7952 &\textbf{4.9804} \\
         \hline
         \multirow{4}{*}{Toys} & HR@5 & 4.0813 & 5.9959 & 4.0246 & 3.9605 & 6.0529 &\textbf{6.2539} \\
          & NDCG@5 & 2.8858 & 4.0643 & 2.5603 & 2.8067 & 4.4397 &\textbf{4.6209} \\
          & HR@10 & 5.6610 & 8.1911 & 5.9105 & 5.5336 & 8.1090 &\textbf{8.2261}\\
          & NDCG@10 & 3.3927 & 4.7712 & 3.1662 & 3.3144 & 5.0672 &\textbf{5.2578} \\
         \hline
    \end{tabular}
    }
\end{table}
% \vspace{-0.35in}

\subsection{Hyperparameter Sensitivity Analysis}
We analyze the impact of two important hyperparameters $\alpha$ and $\beta$, which control the relative weight of the conditional flow matching loss $\mathcal{L}_{\text{CFM}}$ and the single-step alignment loss $\mathcal{L}_{\text{Align}}$, respectively. The results on the Toys and Beauty datasets are shown in Figure~\ref{Hyperparameter}. When $\alpha$ increases from small values, performance improves steadily, indicating that moderate trajectory-level supervision facilitates the learning of smooth preference flow. However, excessive $\alpha$ weakens the influence of $\mathcal{L}_{\text{Align}}$, leading to performance degradation. A similar pattern is observed for $\beta$: proper alignment regularization enhances ranking discrimination.

Overall, FlowRec maintains stable performance across a wide range of hyperparameter values, demonstrating its robustness to the balance between trajectory guidance and alignment supervision.

\begin{figure}[H]
    \centering
    \includegraphics[width=\columnwidth]{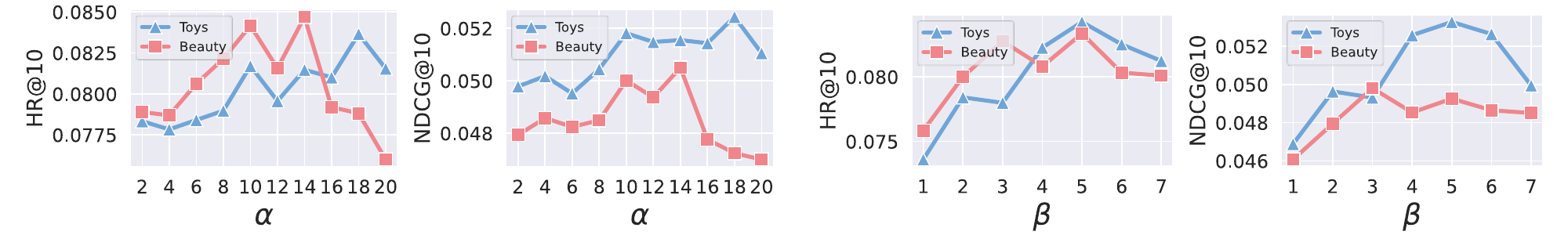}
    \caption{The sensitivity of FlowRec to $\alpha$ and $\beta$.}
    \label{Hyperparameter}
\end{figure}
\vspace{-0.45in}

\subsection{Inference Efficiency Analysis}
\begin{wrapfigure}{r}{0.5\linewidth}
    \vspace{-0.15in}
  \centering
  \includegraphics[width=\linewidth]{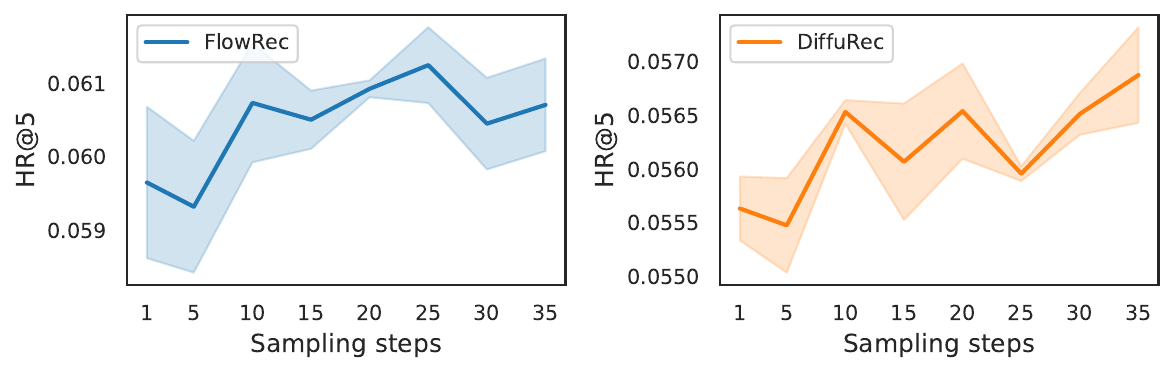}
  \caption{The impact of sampling steps on FlowRec and DiffuRec.}
  \label{efficiency}
  \vspace{-0.15in}
\end{wrapfigure}
\subsubsection{Sampling Steps.} The number of sampling steps is a key factor influencing the inference speed. To evaluate FlowRec's inference efficiency, we varied the number of sampling steps on the Toys dataset. As illustrated in Figure~\ref{efficiency}, FlowRec achieves optimal performance with only 8-10 sampling steps, whereas DiffuRec requires 30-35 steps. Even with a single step, FlowRec remains competitive, benefiting from its straight flow formulation and the simple, efficient ODE solver that avoids iterative denoising.

\vspace{-0.1in}
\subsubsection{Inference Time.}  We further compare the training and inference time of different models on the Toys and Yelp datasets, as summarized in Table~\ref{tab:infer_time}. For fairness, both DiffuRec and FlowRec are evaluated with 10 sampling steps. FlowRec exhibits comparable training efficiency to DiffuRec but achieves substantially faster inference, with runtime close to SASRec. This improvement arises from the use of a simple Euler ODE solver, which less computationally intensive than DiffuRec's stochastic denoising process.

\begin{table}
    \centering
    \caption{Training and inference time (s/epoch).}
    \label{tab:infer_time}
    \renewcommand{\arraystretch}{1}
    \tabcolsep=0.3cm
    \resizebox{\columnwidth}{!}{
        \begin{tabular}{ccccccc}
            \hline
             Datasets & Phase & SASRec & ACVAE & PreferDiff & DiffuRec & FlowRec  \\
             \hline
             \multirow{2}{*}{Toys} & Training & 5.33 & 8.687 & 31.99 & 10.11 & 9.98 \\
              & Inference & 0.88 & 18.00 & 5.25 & 8.51 & 1.27 \\
             \hline
             \multirow{2}{*}{Yelp} & Training  & 11.67 & 13.59 & 58.18 & 19.10 & 17.62 \\
              & Inference  & 2.93 & 28.99 & 8.30 & 14.77 & 3.67 \\
             \hline
        \end{tabular}
    }
\end{table}

\section{Conclusion}
 \label{conclution}
This paper introduced FlowRec, a simple yet effective framework for SR  based on the flow matching paradigm. FlowRec explicitly models user preference trajectories from personalized behavior-based priors to target items, enabling stable and efficient preference transitions. By learning a vector field that captures the directional dynamics of user interests and introducing a single-step alignment loss to enforce ranking-aware supervision, FlowRec aligns the generative process with the recommendation objective. Extensive experiments on four benchmark datasets demonstrate that FlowRec consistently surpasses state-of-the-art methods in terms of recommendation accuracy and inference efficiency.

\subsubsection{\ackname} This research was supported by grants from the National Natural Science Foundation of China (No. 62502486), the grants of Provincial Natural Science Foundation of Anhui Province (No. 2408085QF193), USTC ResearchFunds of the DoubleFirst-Class Initiative (No. YD2150002501), the Fundamental Research Funds for the Central Universities of China (No. WK2150110032).

\bibliographystyle{splncs04}
\bibliography{ref.bib}




% LNCS Homepage, \url{http://www.springer.com/lncs}, last accessed 2023/10/25
% \end{thebibliography}
\end{document}